\documentclass[10 pt]{article}
\usepackage{latexsym}
\usepackage{amsmath}
\usepackage{epsfig}

\begin{document}

\baselineskip = 18pt

\author{
Arif Akhundov $^{a, b}$ and Anwar Shiekh $^{c}$\\
\vspace{.5 cm}\\
$^a$ Departamento de F\'isica Te\'orica and IFIC, Universidad de Valencia-CSIC,\\
       E-46100 Burjassot (Valencia), Spain\\
$^b$ Institute of Physics, Azerbaijan Academy of Sciences,\\
       H. Cavid ave. 33, 370143 Baku, Azerbaijan\\
$^c$ Mathematics, Science and Technology division, \\
Din\'e College, Tsaile, AZ 86556, USA
}
\title{A Review of Leading Quantum Gravitational Corrections to Newtonian Gravity}

\maketitle

\begin{abstract}
In this review we present the theoretical background for
treating General Relativity as an effective field theory and
focus on the concrete results of such a treatment. As a result
we present the calculations of the low-energy leading
gravitational corrections to the Newtonian potential between two sources.
\end{abstract}




\section{Introduction}

The fundamental equation of the non-relativistic theory of gravity
is the Newtonian gravitational law, which
predicts the potential energy of the gravitational attraction
between two bodies as:
\begin{equation}
V(r) = -G \frac{m_1 m_2}{r}
\end{equation}
Here $V(r)$ is a measure for the potential energy,
$m_1$ and $m_2$ are the masses of the
two particles, $r$ is the distance between the masses
and $G$ is the universal gravitational constant.

In  contrast the theory of General Relativity~\cite{Einstein} provides a
framework for extending Newton's theory to objects with
relativistic velocities. In general relativity one solves
the basic field equation:
\begin{equation}
R_{\mu\nu}(g_{\mu\nu}) - \frac12 R (g_{\mu\nu}) g_{\mu\nu} = 16\pi
G T_{\mu\nu} - \Lambda g_{\mu\nu}
\end{equation}
where $g_{\mu\nu}$ is the gravitational metric, $R^{\alpha}_{\beta\mu\nu}$ is
the tensor for the curvature of space-time\footnote{\footnotesize{$\left(R_{\mu\nu} = R^\beta_{\
\mu\nu\beta}\right)$, $\left(R \equiv g^{\mu\nu}R_{\mu\nu}\right)$}} and $T_{\mu\nu}$
is the total energy-momentum tensor.
The cosmological constant $\Lambda$ may be needed on cosmological
scales, and is today believed to have a non-zero expectation value
in the Universe. When we solve the Einstein
equation we find the metric which is a local object
that depends on the geometry of space-time. In this way a solution of the
gravitational problem is found. Einstein's description
holds in the fully relativistic regime, and its low-energy and
non-relativistic predictions match the expectations of Newtonian
mechanics.

A longstanding puzzle in Modern Physics is how to wed General
Relativity with the quantum theory. It is not at all obvious how
this might be achieved since General Relativity and quantum
mechanics seem to be based on completely different perceptions of
physics -- nevertheless this question is one of the most pressing
questions of modern theoretical physics and has been the subject of
many studies, $e.g.$, see refs.~\cite{Dewitt},
~\cite{Gupta},~\cite{Faddeev},~\cite{Mandelstam},~\cite{Schwinger1,Schwinger2},~\cite{Weinberg},
~\cite{Iwasaki},~\cite{Veltman1,Veltman2}.

All sorts of interpretational complications arise when trying to quantize
General Relativity. A possible starting point
for such a theory appears to be to interpret General Relativity as a quantum
field theory, to let the metric be the basic gravitational field,
and to quantize the Einstein-Hilbert action:
\begin{equation}
{\cal S }_{\text EH} = \int d^4 x \sqrt{-g}{R\over 16\pi G}
\end{equation}
where $g = \det (g_{\mu\nu})$ and $R$ is the scalar curvature.
However the above action is not self contained under renormalization
since loop diagrams will generate new terms not present in the
original action refs.~\cite{Veltman1,Veltman2,Goroff,vandeVen}. This
is the renowned renormalization problem that hinders the
quantization of general relativity.

One of the physically interesting problem is the calculation of the
leading order quantum corrections to the Newtonian potential which
has been in the focus of many studies in different schemes, using
Feynman diagrams for the loops in the graviton
propagator~\cite{Rad,Capp1,Duff1,Capp2,Capp3,Duff2,Veltman1},
renormalizable $R^2$ gravity~\cite{Odin1,Odin2,Odin3} and
Semiclassical Gravity~\cite{Mazz1,Mazz2,Mazz3,Paul}.

After introducing an effective field theory for processes with a
typical energy less the Planck mass, i.e. with $|q^2|\ll
M_{P}^2\simeq 10^{38}$ GeV$^2$, by Weinberg \cite{Weinberg1}, the
effective theory for gravity can been modeled in a manner analogous
with that of Chiral Perturbation Theory~\cite{Weinberg2} for QCD.
This way to look at General Relativity was proposed by
Shiekh~\cite{Shiekh1} and Donoghue~\cite{Donoghue1}, and they have
shown that reliable quantum predictions at the low energies can be
made.

In spite of fact that unmodified General Relativity is
not renormalizable, be it pure General Relativity or General
Relativity coupled to bosonic or fermionic matter, see
e.g.~\cite{Veltman1,Veltman2,Deser,Gast}, using the framework of an effective field
theory, these theories do become order by order renormalizable in the
low energy limit. When General Relativity is
treated as an effective theory, renormalizability simply fails to be
an issue. The ultraviolet divergences arising e.g. at the 1-loop level
are dealt with by renormalizing the parameters of higher derivative
terms in the action.

When approaching general relativity in this manner, it is convenient
to use the background field method~\cite{Dewitt,Abbott}. Divergent
terms are absorbed away into phenomenological constants which
characterize the effective action of the theory. The price paid is the
introduction of a set of never-ending higher order derivative
couplings into the theory, unless
using the approach of Shiekh~\cite{Shiekh1}.
The effective action contains all terms consistent
with the underlying symmetries of the theory. Perturbatively
only a finite number of terms in the action are required for each
loop order.

In pioneering papers~\cite{Donoghue1} Donoghue first has shown how to derive the
leading quantum and classical relativistic corrections to the Newtonian
potential of two masses. This calculation has since been the focus of a number of
publications~\cite{Muzinich,Hamber,Akh,Khr1,Bohr1,Khr2}, and
this work continues, most recently in the paper~\cite{Kir1}.

Unfortunately, due to difficulty of the calculation and its myriad
of tensor indices there has been some disagreement among the results of
various authors.
The classical component of the corrections were found long ago by
Einstein, Infeld and Hoffmann \cite{Ein}, and by Eddington and
Clark \cite{Edd}. Later this result was reproduced by Iwasaki~\cite{Iwasaki}
by means of Feynman diagrams and has been discussed in the papers
~\cite{Okamura,Barker:bx,Barker:ae}, and here there
is general agreement although there exists an unavoidable ambiguity in defining the potential.

An interesting calculation has been made involving quantum gravitational
corrections to the Schwarzshild and Kerr metrics of
scalars and fermions~\cite{Bohr2,Kir2} where it is shown in detail how
the higher order gravitational contributions to these metrics
emerge from loop calculations. In the papers~\cite{Bohr} and~\cite{Butt}
have been calculated the leading post-Newtonian and quantum corrections to the
non-relativistic scattering amplitude of charged scalars and
spin-{\mbox{\small$\frac{1}{2}$}} fermions in the combined theory of
general relativity and QED. For the recent reviews of
general relativity as an effective field theory, see refs.~\cite{Burgess, Holstein}

Our notations and conventions on the metric tensor, the gauge-fixed
gravitational action, etc. are the same as in \cite{Akh}, namely
($\hbar = c =1$) as well as the Minkowski metric convention $(+1,-1,-1,-1)$.

\section{The quantization of General Relativity}

The Einstein action for General Relativity has the form:
\begin{equation}
{\cal S} = \int d^4x \sqrt{-g}\left[\frac{2R}{\kappa^2}+ {\cal L}_{\text{matter}}\right]
\end{equation}
where $\kappa^2 = 32\pi G$ is defined as the gravitational
coupling, and the curvature tensor is defined as:
\begin{equation}
R^\mu_{\nu\alpha\beta} \equiv
\partial_\alpha \Gamma_{\nu\beta}^\mu
-\partial_\beta \Gamma_{\nu\alpha}^\mu
+\Gamma^\mu_{\sigma\alpha}\Gamma_{\nu\beta}^\sigma
-\Gamma_{\sigma\beta}^\mu\Gamma_{\nu\alpha}^\sigma
\end{equation}
and
\begin{equation}
\Gamma^{\lambda}_{\alpha \beta} =  {1 \over 2} g^{\lambda \sigma}
\left(\partial_{\alpha} g_{\beta \sigma} + \partial_{\beta} g_{\alpha \sigma} -
\partial_{\sigma} g_{\alpha \beta} \right)
\end{equation}
The term $\sqrt{-g}{\cal L}_{\rm matter}$ is a covariant expression
for the inclusion of matter into the theory. We can include any type
of matter. As a classical theory the above Lagrangian defines the
theory of general relativity.

Massive spinless matter fields interact with the gravitational field as
described by the action

\begin{equation}
{\cal S}_{matter} = \int d^4x \sqrt{-g} \left[ {1 \over 2} g^{\mu \nu}
\partial_{\mu} \phi \partial_{\nu} \phi - {1 \over 2} m^2 \phi^2 \right]
\end{equation}

Any effective field theory can be seen as an expansion in energies
of the light fields of the theory below a certain scale. Above the
scale transition energy there will be additional heavy fields that
will manifest themselves. Below the transition the heavy degrees
of freedom will be integrated out and will hence not contribute to
the physics. Any effective field theory is built up from terms
with higher and higher numbers of derivative couplings on the
light fields and obeying the gauge symmetries of the basic theory.
This gives us a precise description of how to construct effective
Lagrangians from the gauge invariants of the theory. We expand the
effective Lagrangian in the invariants ordered in magnitude of
their derivative contributions.

An effective treatment of pure General Relativity results in the following Lagrangian:
\begin{equation}
{\cal L}_{grav} =
\sqrt{-g}\left[\frac{2R}{\kappa^2}+c_{1}R^2+c_{2}R^{\mu\nu}R_{\mu\nu}+\ldots\right]
\end{equation}
where the ellipses denote that the effective action is in fact an infinite
series---at each new loop order additional higher derivative terms must be
taken into account. This Lagrangian includes all possible higher derivative
couplings, and every coupling constant in the Lagrangian is considered to be determined
empirically unless set to zero to achieve causality~\cite{Shiekh1}. Similarly one must include higher derivative contributions to
the matter Lagrangian in order to treat this piece of the Lagrangian as an
effective field theory~\cite{Donoghue1}.

Computing the leading low-energy quantum
corrections of an effective field theory, a useful distinction is
between non-analytical and analytical contributions from the
diagrams. Non-analytical contributions are generated by
the propagation of two or more massless particles in the Feynman diagrams.
Such non-analytical effects are long-ranged and, in the low energy limit of the
effective field theory, they dominate over the analytical
contributions which arise from the propagation of massive particles.
The difference between massive and massless
particle modes originates from the impossibility of expanding a
massless propagator $\sim 1/{q^2}$ while:
\begin{equation}
\frac{1}{q^2-m^2} = -\frac{1}{m^2}\Big(1+\frac{q^2}{m^2}+\ldots \Big)
\end{equation}
No $1/{q^2}$ terms are generated in the above expansion of the
massive propagator, thus such terms all arise from the propagation
of massless modes. The analytical contributions from the diagrams are local effects and
thus expandable in power series.

Non-analytical effects are typically originating from terms which in the
S-matrix go as, $e.g.$, $\sim \ln(-q^2)$ or
$\sim 1/\sqrt{-q^2}$, while the generic example of an analytical contribution
is a power series in momentum $q$. Our interest is
only in the non-local effects, thus we will only consider the
non-analytical contributions of the diagrams.

The procedure of the background field quantization is as follows.
The quantum fluctuations of the gravitational field are expanded about
a smooth background metric $\bar g_{\mu\nu}$~\cite{Veltman1,Veltman2}, i.e. flat space-time
$\bar g_{\mu\nu}\equiv\eta_{\mu\nu} = diag(1,-1,-1,-1)$, and the metric $g_{\mu\nu}$ is the sum of this
background part  and a quantum contribution $\kappa h_{\mu\nu}$:
\begin{equation}
g_{\mu\nu}\equiv \bar g_{\mu\nu} + \kappa h_{\mu\nu}
\end{equation}
From this equation we get the expansions for the upper metric field $g^{\mu\nu}$,
and for $\sqrt{-g}$:
\begin{equation}\begin{split}
g^{\mu\nu} &= \bar g^{\mu\nu} - \kappa h^{\mu\nu} + \ldots\\
\sqrt{-g} &= \sqrt{-\bar g}\left[1+ \frac12\kappa h + \ldots\right]
\end{split}\end{equation}
where $h^{\mu\nu}\equiv \bar g^{\mu\alpha} \bar g^{\nu\beta} h_{\alpha\beta}$
and $ h  \equiv \bar g^{\mu\nu} h_{\mu\nu}$.

The corresponding curvatures are given by
\begin{eqnarray}
\bar R_{\mu\nu}&=&{\kappa\over 2}\left[\partial_\mu\partial_\nu h
+\partial_\lambda\partial^\lambda h_{\mu\nu}
-\partial_\mu\partial_\lambda {h}^\lambda{}_\nu
-\partial_\nu\partial_\lambda {h}^\lambda{}_\mu\right]\nonumber\\
\bar R &=&\bar g^{\mu\nu} \bar R_{\mu\nu}=\kappa\left[\Box h
-\partial_\mu\partial_\nu h^{\mu\nu}\right]
\end{eqnarray}

In order to quantize the field $ h_{\mu\nu}$ one needs to fix the gauge.
In the harmonic (or deDonder) gauge~\cite{Veltman1} ---$g^{\mu\nu} \Gamma^\lambda_{\mu\nu}=0$---which requires, to first order in the field expansion,
\begin{equation}
\partial^\beta h_{\alpha\beta}-{1\over 2}\partial_\alpha h = 0
\end{equation}

In the quantization, the Lagrangians are expanded in the gravitational
fields, separated in quantum and background parts, and the vertex factors
as well as the propagator are derived from the expanded action.

The expansion of the Einstein action takes the form~\cite{Veltman1,Veltman2}:

\begin{equation}
{\cal S}_{grav} = \int d^4x \sqrt{-\bar{g}} \left[ {2 \bar{R} \over
\kappa^2} +
{\cal L}^{(1)}_g + {\cal L}^{(2)}_g + \ldots \right]
\end{equation}

\noindent where the subscripts count the number of powers of $\kappa$ and

\begin{eqnarray}
{\cal L}^{(1)}_g & = & {h_{\mu \nu} \over \kappa} \left[ \bar{g}^{\mu
\nu} \bar{R} -
2 \bar{R}^{\mu \nu} \right] \nonumber \\
{\cal L}^{(2)}_g & = &  {1 \over 2} {D_\alpha} h_{\mu \nu} {D^\alpha} h^{\mu \nu}
 - {1 \over 2} {D_\alpha} h {D^\alpha} h + {D_\alpha} h {D_\beta} h^{\alpha \beta}
 - {D_\alpha} h_{\mu \beta} {D^\beta} h^{\mu \alpha} \nonumber \\
& & + \bar{R} \left( {1 \over 4} h^{2} - {1 \over 2} h_{\mu \nu} h^{\mu
\nu} \right) + \left( 2 h^{\lambda}_{\mu} h_{\nu \lambda} - h h_{\mu \nu}
\right)
\bar{R}^{\mu \nu}
\end{eqnarray}
\noindent where $D_\alpha$ denotes the covariant derivative with respect to the background metric.

\noindent   A similar expansion of the matter action yields~\cite{Donoghue1}:

\begin{equation}
{\cal S}_{matter} = \int d^4x \sqrt{-\bar{g}} \left [ {\cal L}^0_m + {\cal L}^{(1)}_m
+ {\cal L}^{(2)}_m + \ldots \right]
\end{equation}

\noindent with

\begin{eqnarray}
{\cal L}^{(0)}_m & = & {1 \over 2} \left( \partial_{\mu} \phi
\partial^{\mu} \phi
- m^2 \phi^2 \right) \nonumber \\
{\cal L}^{(1)}_m & = & - {\kappa \over 2} h_{\mu \nu} T^{\mu \nu}
\nonumber
\\
T_{\mu \nu} & \equiv &  \partial_{\mu} \phi \partial_{\nu} \phi - {1 \over
2}
\bar{g}_{\mu \nu} \left( \partial_{\lambda} \phi \partial^{\lambda} \phi -
m^2
\phi^2 \right) \nonumber \\
{\cal L}^{(2)}_m & = &  \kappa{^2} \left( {1 \over 2} h^{\mu \nu}
h^{\nu}_{\lambda}
- {1 \over 4} h h^{\mu \nu} \right) \partial_{\mu} \phi \partial_{\nu} \phi
\nonumber \\
& & - {\kappa{^2} \over 8} \left( h^{\lambda \sigma} h_{\lambda \sigma}
- {1 \over 2} h h
\right)
\left[
\partial_{\mu} \phi \partial^{\mu} \phi - m^2 \phi^2 \right]
\end{eqnarray}

The background metric $\bar R^{\mu\nu}$ should satisfy Einstein's equation
\begin{equation}
\bar{R}^{\mu \nu} - {1 \over 2} \bar{g}^{\mu \nu} \bar{R} = {\kappa^2\over 4} T^{\mu \nu}
\end{equation}
and the linear terms in $h_{\mu \nu}, {\cal L}^{(1)}_g + {\cal L}^{(1)}_m$, is vanishing.

For the calculation of the quantum gravitational corrections at one loop, we need to consider
the following actions:
\begin{eqnarray}
{\cal S}_0 & = & \int d^4 x \sqrt{-g} \left [ { 2 {\bar R} \over \kappa^2} + {\cal L}^{(0)}_m \right ]
\nonumber \\
{\cal S}_{2} & = & \int d^4 x \sqrt{-g} \left [ {\cal L}^{(2)}_g + {\cal L}^{(2)}_m
                    + {\cal L}_{gauge} + {\cal L}_{ghost}  \right ]
\end{eqnarray}

\noindent with the gauge fixing Lagrangian~\cite{Veltman1}

\begin{equation}
{\cal L}_{gauge} =  \left ( D^{\nu} h_{\mu \nu} - {1 \over 2} D_{\mu} h \right)
                 \left( D_{\lambda} h^{\mu \lambda} - {1 \over 2} D^{\mu} h \right)
\end{equation}
\noindent and the ghost Lagrangian
\begin{equation}
{\cal L}_{ghost} = \eta^{\ast \mu} \left ( D_{\lambda} D^{\lambda}\eta_{\mu}
                    - \bar{R}_{\mu \nu} \eta^{\nu} \right )
\end{equation}
\noindent for the Faddeev-Popov field $ \eta_{\mu}$.

\section{The Feynman rules}

From the Lagrangians (18) we can derive the list of Feynman rules~\cite{Bohr1}.

$\bullet$ {\bf Scalar propagator}\noindent\\
The massive scalar propagator is:\\ \\

\begin{minipage}[h]{0.4\linewidth}\vspace{0.4cm}
\centering\includegraphics[scale=1]{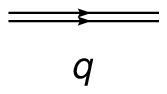}
\end{minipage}
\begin{minipage}[h]{0.65\linewidth}
\centering$\displaystyle = \frac i{q^2-m^2+i\epsilon}$
\end{minipage}

$\bullet$ {\bf Graviton propagator}\noindent\\
The graviton propagator in harmonic gauge is:\\ \\

\begin{minipage}[h]{0.4\linewidth}\vspace{0.4cm}
\centering\includegraphics[scale=1]{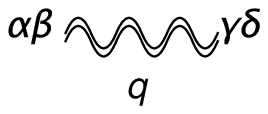}
\end{minipage}
\begin{minipage}[h]{0.65\linewidth}
\centering $\displaystyle = \frac {i{\cal
P}^{\alpha\beta\gamma\delta}}{q^2+i\epsilon}$
\end{minipage}\\
where
 $${\cal P}^{\alpha\beta\gamma\delta} =
 \frac12\left[\eta^{\alpha\gamma}\eta^{\beta\delta}
+ \eta^{\beta\gamma}\eta^{\alpha\delta}
-\eta^{\alpha\beta}\eta^{\gamma\delta}\right]$$

$\bullet$ {\bf 2-scalar-1-graviton vertex}\noindent\\
The two scalar - one graviton vertex is:\\

\begin{minipage}[h]{0.4\linewidth}\vspace{0.15cm}
\centering\includegraphics[scale=1]{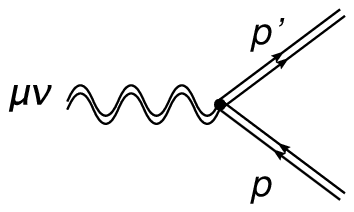}
\end{minipage}
\begin{minipage}[h]{0.65\linewidth}
\centering$\displaystyle = \tau^{\mu \nu}(p,p',m)$
\end{minipage}\vspace{0.3cm}\\
where
$$\tau^{\mu\nu}(p,p',m) = -\frac{i\kappa}2\left[p^\mu p^{\prime \nu}
+p^\nu p^{\prime \mu} - \eta^{\mu\nu}\left((p\cdot
p^\prime)-m^2\right)\right]
$$

$\bullet$ {\bf 2-scalar-2-graviton vertex}\noindent\\
The two scalar - two graviton vertex is \\

\begin{minipage}[h]{0.4\linewidth}\vspace{0.15cm}
\centering\includegraphics[scale=1]{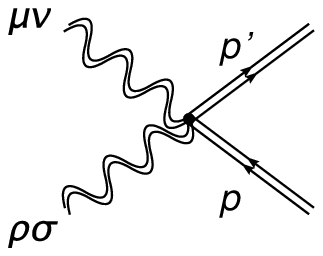}
\end{minipage}
\begin{minipage}[h]{0.65\linewidth}
\centering$\displaystyle = \tau^{\eta\lambda\rho\sigma}(p,p',m)$
\end{minipage}\vspace{0.3cm}
where
\begin{equation}{\begin{aligned}
\tau^{\eta \lambda \rho \sigma}(p,p')&= {i\kappa^2} \bigg [ \left
\{I^{\eta
\lambda\alpha \delta} {I}^{\rho \sigma\beta}_{\ \ \ \ \delta} -
\frac14\left\{\eta^{\eta \lambda} I^{\rho \sigma\alpha \beta} +
 \eta^{\rho \sigma} I^{\eta \lambda\alpha \beta} \right \}
\right \} \left (p_\alpha p^\prime_{\beta} + p^\prime_{\alpha} p_\beta
\right ) \\
&-\frac12 \left \{ I^{\eta \lambda\rho \sigma} - \frac12\eta^{\eta
\lambda}\eta^{\rho \sigma} \right \} \left [ (p\cdot p') - m^2
\right]\bigg]\end{aligned}}
\end{equation}
with
$$I_{\alpha\beta\gamma\delta}={1\over
2}(\eta_{\alpha\gamma}\eta_{\beta\delta}
+\eta_{\alpha\delta}\eta_{\beta\gamma})$$

$\bullet$ {\bf 3-graviton vertex}\noindent\\
The three graviton vertex is:

\begin{minipage}[h]{0.4\linewidth}\vspace{0.15cm}
\centering\includegraphics[scale=1]{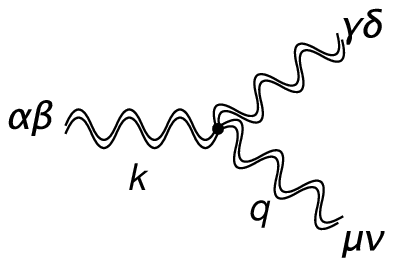}
\end{minipage}
\begin{minipage}[h]{0.65\linewidth}
\centering$\displaystyle =
{\tau}_{\alpha\beta\gamma\delta}^{\mu\nu}(k,q)$
\end{minipage}\vspace{0.3cm}
where
\begin{equation}{
\begin{aligned}
{\tau}_{\alpha  \beta \gamma \delta }^{\mu
\nu}(k,q)&=-\frac{i\kappa}2\times
\bigg({\cal P}_{\alpha \beta \gamma \delta }\bigg[k^\mu k^\nu+ (k-q)^\mu
(k-q)^\nu
+q^\mu q^\nu-
\frac32\eta^{\mu \nu}q^2\bigg]\\[0.00cm]&
+2q_\lambda q_\sigma\bigg[ I_{\alpha \beta }^{\ \ \
\sigma\lambda}I_{\gamma \delta
}^{\ \ \ \mu \nu} + I_{\gamma \delta }^{\ \ \ \sigma\lambda}I_{\alpha
\beta }^{\ \ \
\mu \nu} -I_{\alpha \beta }^{\ \ \ \mu  \sigma} I_{\gamma \delta }^{\ \ \
\nu
\lambda} - I_{\gamma \delta }^{\ \ \ \mu \sigma} I_{\alpha \beta }^{\ \ \
\nu
\lambda}
\bigg]\\[0cm]&
+\bigg[q_\lambda q^\mu \bigg(\eta_{\alpha \beta }I_{\gamma \delta }^{\ \ \
\nu
\lambda}+\eta_{\gamma \delta }I_{\alpha \beta }^{\ \ \ \nu
\lambda}\bigg) +q_\lambda
q^\nu \left(\eta_{\alpha \beta }I_{\gamma \delta }^{\ \ \ \mu
\lambda}+\eta_{\gamma
\delta }I_{\alpha \beta }^{\ \ \ \mu  \lambda}\right)\\&
-q^2\left(\eta_{\alpha
\beta }I_{\gamma \delta }^{\ \ \ \mu \nu}-\eta_{\gamma \delta }I_{\alpha
\beta }^{\
\ \ \mu \nu}\right) -\eta^{\mu \nu}q_\sigma q_\lambda\left(\eta_{\alpha
\beta
}I_{\gamma \delta }^{\ \ \ \sigma\lambda} +\eta_{\gamma \delta }I_{\alpha
\beta }^{\
\ \
\sigma\lambda}\right)\bigg]\\[0cm]&
+\bigg[2q_\lambda\big(I_{\alpha \beta }^{\ \ \ \lambda\sigma}I_{\gamma
\delta
\sigma}^{\ \ \ \ \nu}(k-q)^\mu +I_{\alpha \beta }^{\ \ \
\lambda\sigma}I_{\gamma
\delta \sigma}^{\ \ \ \ \mu }(k-q)^\nu -I_{\gamma \delta }^{\ \ \
\lambda\sigma}I_{\alpha \beta \sigma}^{\ \ \ \ \nu}k^\mu -I_{\gamma \delta
}^{\ \ \
\lambda\sigma}I_{\alpha \beta \sigma}^{\ \ \ \ \mu }k^\nu \big)\\&
+q^2\left(I_{\alpha \beta \sigma}^{\ \ \ \ \mu }I_{\gamma \delta }^{\ \ \
\nu
\sigma} + I_{\alpha \beta }^{\ \ \ \nu \sigma}I_{\gamma \delta \sigma}^{\
\ \ \ \mu
}\right) +\eta^{\mu \nu}q_\sigma q_\lambda\left(I_{\alpha \beta }^{\ \ \
\lambda\rho}I_{\gamma \delta  \rho}^{\ \ \ \ \sigma} +I_{\gamma \delta
}^{\ \ \
\lambda\rho}I_{\alpha \beta  \rho}^{\ \ \ \
\sigma}\right)\bigg]\\[0cm]&
+\bigg\{(k^2+(k-q)^2)\big[I_{\alpha \beta }^{\ \ \ \mu  \sigma}I_{\gamma
\delta
\sigma}^{\ \ \ \ \nu} +I_{\gamma \delta }^{\ \ \ \mu  \sigma}I_{\alpha
\beta
\sigma}^{\ \ \ \ \nu} -\frac12\eta^{\mu \nu}{\cal P}_{\alpha \beta \gamma
\delta
}\big]\\&-\left(I_{\gamma \delta }^{\ \ \ \mu \nu}\eta_{\alpha \beta
}k^2+I_{\alpha
\beta }^{\ \ \ \mu \nu}\eta_{\gamma \delta }(k-q)^2\right)\bigg\}\bigg)
\end{aligned}}
\end{equation}

\section{Scattering amplitude and potential}

The general form for any diagram contributing to the scattering amplitude
of gravitational interactions of two masses is:
\begin{equation}
{\cal M} \sim
\Big(A+Bq^2+\ldots+{C_0}\kappa^4\frac1{q^2}+{C_1}\kappa^4\ln(-q^2)+
{C_2}\kappa^4\frac{m}{\sqrt{-q^2}} +\ldots \Big)
\end{equation}
where $A, B, \ldots$ correspond to the local analytical interactions
which are of no interest to us (these terms will only dominate
in the high energy regime of the effective theory) and
$C_0, C_1, C_2, \ldots$ correspond to the
non-local, non-analytical interactions.

The $C_1$ and $C_2$ terms will yield the leading quantum gravitational and relativistic
post-Newtonian corrections to the Newtonian potential.
The space parts of the non-analytical terms Fourier transform as:
\begin{equation}\begin{split}
\int \frac {d^3 q}{(2\pi)^3} ~e^{i{\bf q}\cdot {\bf r}}
\frac 1{|{\bf q}|^2}& = \frac {1}{4\pi r}\\
\int \frac {d^3 q}{(2\pi)^3} ~ e^{i{\bf q}\cdot {\bf r}}
\frac 1{|{\bf q}|}& = \frac {1}{2\pi^2 r^2}\\
\int \frac {d^3 q}{(2\pi)^3} ~e^{i{\bf q}\cdot {\bf r}}
\ln({\bf q}^2)& = \frac {-1}{2\pi r^3}\\
\end{split}\end{equation}
so clearly these terms will contribute to the corrections.

The importance of these transforms, is that they are
from non-analytic terms in momentum space and so cannot
be renormalized into the original Lagrangian, and as such
one might anticipate that they are of finite magnitude.
Because of this, the problem of renormalizing quantum gravity is put off.

In the quantization of General Relativity the definition of a
potential is certainly not obvious. One can choose between several
definitions of the potential depending on, $e.g.$, the physical
situation, how to define the energy of the fields, the diagrams
included etc. The choice of potential, which includes all 1-loop diagrams~\cite{Hamber,Kazakov},
is the simplest, gauge invariant definition of the potential.

The calculation of the non-relativistic potential using the the
full amplitude is as follows. First, to relate the expectation value for
the $S$ matrix to the Fourier transform of the potential
$\tilde V({\bf q})$ in the non-relativistic limit:

\begin{equation}
\langle k_1, k_3 | S | k_2, k_4 \rangle = -i\tilde {V}({\bf q})(2\pi)\delta(E_{i} - E_{f})
\end{equation}
where $k_1$, $k_3$ and $k_2$, $k_4$ are the incoming and
outgoing momentum respectively, $q \equiv k_2-k_1 = k_3-k_4$, and $E_i-E_f$ is the energy
difference between the incoming and outgoing states.
The invariant matrix element $i{\cal M}$ resulting from the diagrams is:
\begin{equation}
\langle k_1, k_3 | S | k_2, k_4 \rangle = (2\pi)^4\delta^{(4)}(k_2+k_4-k_1-k_3)(i{\cal M})
\end{equation}
In the non-relativistic limit ($q=(0,\bf q)$) we have:
\begin{equation}
\tilde V({\bf q}) = -\frac{1}{2m_1}\frac{1}{2m_2}{\cal M}
\end{equation}
so that
\begin{equation}
V({\bf x}) = -\frac{1}{2m_1}\frac{1}{2m_2}\int
\frac{d^3k}{(2\pi)^3}e^{i{\bf k} \cdot {\bf x}}{\cal M}
\end{equation}

This is how we define the non-relativistic potential V({\bf q})
generated by the considered non-analytic parts. In the above equation
${\cal M}$ is the non-analytical part of the amplitude of the scattering
process in non-relativistic limit to a given loop order~\cite{Hamber}.

\section{The contributions of Feynman diagrams}

In general, the Feynman rules are large and the tensor
algebra immense. Much of the calculational simplicity should be
restored by submitting this part of the complexity to the
computer. However, the intermediate results can be so
extensive that even a super-computer can choke without help.
For example, imagine one had the contraction of three tensors:
$
\alpha^{\mu \nu} \beta^{\rho \sigma}
\gamma_{\mu \nu \rho \sigma}
$
each of which consists of many terms. Then the
computer, in trying to contract out the indices, tends to
expand out the entire expression, which can easily lead to
thousands of terms in the intermediate expression, and so
overpower the computers memory. The resolution lies in asking
the computer to initially expand out only $\alpha$ for example:
$
(\alpha_1^{\mu \nu} + \alpha_2^{\mu \nu} + \ldots)
\beta^{\rho \sigma} \gamma_{\mu \nu \rho \sigma}
$.
In this way the computer is presented with several terms that
can each be contracted separately. This seemingly innocuous
move can make all the difference between the machine being
able to perform the calculation or not. It is fine details
like this that in practice can occupy much of the investigators
time.

The best way to perform such kind of tensor algebra on a computer is use the
Ricci package \cite{Lee} under the Mathematica \cite{Math} program.

\subsection{Vacuum polarization}

't Hooft and Veltman~\cite{{Veltman1}} were the first to calculate
the vacuum polarization diagram in gravity. For the contribution of the graviton plus ghost
vacuum polarization Feynman diagrams\footnote{The Feynman graphs have been plotted with {\bf JaxoDraw}
~\cite{JaxoDraw}}:

\begin{figure}[ht]
\begin{minipage}[t]{0.46\linewidth}
\begin{center}
\includegraphics[scale=1]{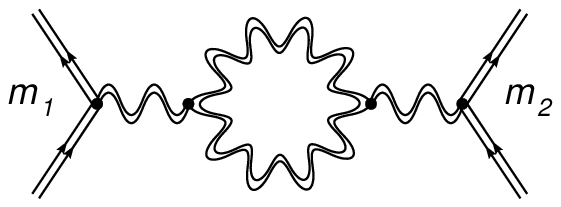}
\end{center}
\end{minipage}
\begin{minipage}{0.1\linewidth}\ \end{minipage}
\begin{minipage}[t]{0.46\linewidth}
\begin{center}
\includegraphics[scale=1]{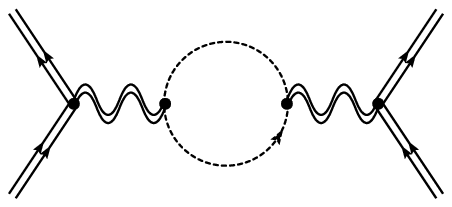}
\end{center}
\end{minipage}
\end{figure}
\noindent we have:
\begin{eqnarray}
 {\Pi}_{\alpha\beta\gamma\delta}&=&-{\textstyle{\kappa ^2} \over \textstyle{16 \pi^2}}
 L \left[{21\over 120}q^4
I_{\alpha\beta\gamma\delta}+{23\over
120}q^4\eta_{\alpha\beta}\eta_{\gamma\delta} \right.\nonumber
-\left.{23\over 120}q^2(\eta_{\alpha\beta}q_\gamma q_\delta
+\eta_{\gamma\delta}q_\alpha
q_\beta)\right.\\
&-&\left.{21\over 240}q^2(q_\alpha q_\delta\eta_{\beta\gamma}+q_\beta
q_\delta\eta_{\alpha\gamma} +q_\alpha
q_\gamma\eta_{\beta\delta}+q_\beta q_\gamma\eta_{\alpha\delta})
+{11\over 30}q_\alpha q_\beta q_\gamma
q_\delta\right]
\end{eqnarray}
\noindent where $L \equiv \log(-q^2) = \ln({\bf q}^2)$

The result after contracting the various indices is~\cite{Donoghue1,Akh,Bohr1,Khr1}:
\begin{equation}
{\cal M}_{vac}({\bf q})={43\over 15}G^2 m_1 m_2 L
\end{equation}
\noindent The Fourier transform gives the following contribution to the scattering potential
\begin{equation}
V_{vac}(r) = -\frac{43}{30\pi}G^2\frac{m_1m_2}{r^3}
\end{equation}

\subsection{Double-seagull contribution}
The calculation of the double-seagull loop diagram:\\
\begin{figure}[h]
\begin{center}
\includegraphics[scale=1]{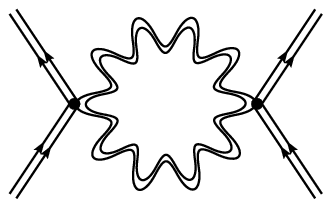}
\end{center}
\end{figure}

\noindent is quite straightforward.
The resulting amplitude is
\begin{equation}
{\cal M}_{seag}({\bf q})=44G^2m_1m_2 L
\end{equation}
\noindent whose Fourier transform yields the double-seagull contribution to the
potential \cite{Khr1, Bohr1}:
\begin{equation}
V_{seag}(r) = -\frac{22}{\pi}G^2\frac{m_1 m_2}{r^3}
\end{equation}

\subsection{The triangle diagrams}
The calculation of the triangle loop diagram:

\begin{figure}[h]
\begin{minipage}[t]{0.46\linewidth}
\begin{center}
\includegraphics[scale=1]{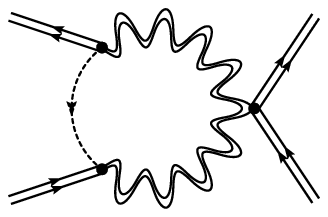}
\end{center}
{\begin{center}(a)\end{center}}
\end{minipage}
\begin{minipage}[t]{0.46\linewidth}
\begin{center}
\includegraphics[scale=1]{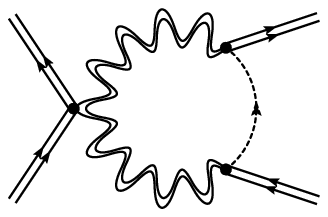}
\end{center}
{\begin{center}(b)\end{center}}
\end{minipage}
\end{figure}
\noindent yields no real complications:
\begin{eqnarray}
{\cal M}_{tri}^{a}(\bf q)&=&-8G^2m_1m_2\left({7\over 2} L +{\pi^2m_1\over |\bf q|}\right)\nonumber\\
{\cal M}_{tri}^{b}(\bf q)&=&-8G^2m_1m_2\left({7\over 2} L +{\pi^2m_2\over |\bf q|}\right)
\end{eqnarray}
\noindent and the Fourier transformed result is~\cite{Donoghue1,Akh,Khr1,Bohr1}:
\begin{equation}
V_{tri}(r) = -4 G^2\frac{m_1m_2(m_1+m_2)}{r^2}+ \frac{28}{\pi}G^2\frac{m_1m_2}{r^3}
\end{equation}

\subsection{Vertex corrections}

Two classes of diagrams go into the set of vertex corrections.
There are two diagrams with a massless graviton in the loop:\\

\begin{figure}[h]
\begin{minipage}[t]{0.46\linewidth}
\begin{center}
\includegraphics[scale=1]{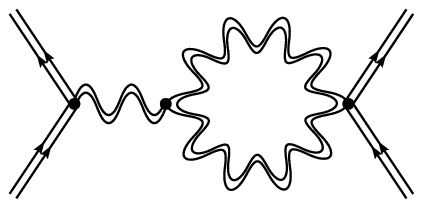}
\end{center}
{\begin{center}(a)\end{center}}
\end{minipage}
\begin{minipage}{0.1\linewidth}\ \end{minipage}
\begin{minipage}[t]{0.46\linewidth}
\begin{center}
\includegraphics[scale=1]{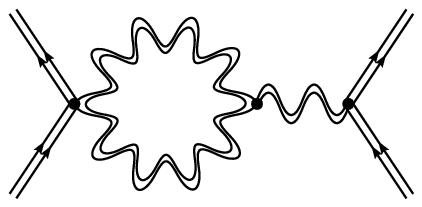}
\end{center}
{\begin{center}(a)\end{center}}
\end{minipage}
\end {figure}
\noindent The calculation of these diagrams is sufficiently simple and results in:
\begin{eqnarray}
{\cal M}_{vert}^{a}(\bf q)&=&-{52\over 3}G^2m_1m_2 L
\end{eqnarray}

\noindent Much more tedious is the calculation of the vertex diagrams with massive particle in the loop:\\
\begin{figure}[h]
\begin{minipage}[t]{0.46\linewidth}
\begin{center}
\includegraphics[scale=1]{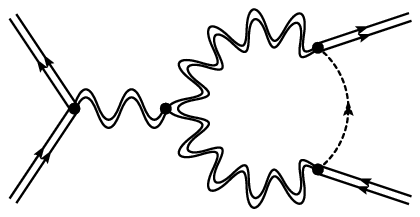}
\end{center}
{\begin{center}(b)\end{center}}
\end{minipage}
\begin{minipage}{0.1\linewidth}\ \end{minipage}
\begin{minipage}[t]{0.46\linewidth}
\begin{center}
\includegraphics[scale=1.1]{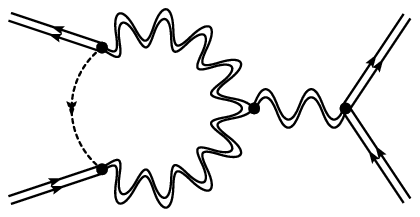}
\end{center}
{\begin{center}(b)\end{center}}
\end{minipage}
\end{figure}

\noindent The result is:
\begin{eqnarray}
{\cal M}_{vert}^{a}(\bf q)&=& 2G^2m_1m_2 \left( {5\over 3}L + {\pi^2(m_1+m_2)\over |\bf q|}\right)
\end{eqnarray}

The vertex diagrams are among the most complicated to calculate. The
first results for these diagrams date back to the original
calculation of Donoghue~\cite{Donoghue1} --- but because
of an algebraic error in the calculation, the original result was
in error and despite various checks of the
calculation~\cite{Akh,Khr1} the correct result has not been given
until~\cite{Bohr1}.

The Fourier transform yields the following result for the
vertex modification of the scattering potential~\cite{Bohr1,Khr2}:
\begin{equation}
V_{vert}^{a}(r) = \frac{26}{3\pi} G^2 \frac{m_1m_2}{r^3}
\end{equation}
and
\begin{equation}
V_{vert}^{b}(r) = G^2\frac{m_1m_2(m_1+m_2)}{r^2}-\frac{5}{3\pi}G^2\frac{m_1m_2}{r^3}
\end{equation}

\subsection{The box diagrams}
The contribution of the box and crossed box diagrams:\\

\begin{figure}[h]
\begin{minipage}[t]{0.46\linewidth}
\begin{center}
\includegraphics[scale=1]{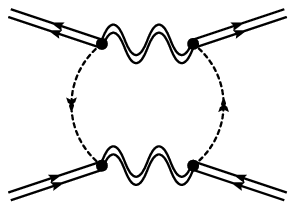}
\end{center}
\end{minipage}
\begin{minipage}[t]{0.46\linewidth}
\begin{center}
\includegraphics[scale=1]{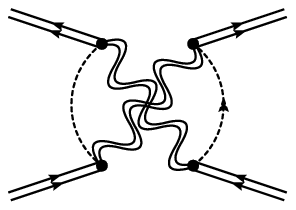}
\end{center}
\end{minipage}

\end{figure}

\noindent to the scattering amplitude in the non-relativistic limit is:
\begin{equation}
{\cal M}_{box}({\bf q})={94\over 3} G^2 m_1m_2 L
\end{equation}
and to the potential~\cite{Khr1,Bohr1}:
\begin{equation}
V_{box}(r) = -\frac{47}{3\pi}G^2\frac{m_1m_2}{r^3}
\end{equation}

\section{The gravitational corrections}

Adding up all one-loop gravitational corrections we
have the final result for the non-relativistic Newtonian potential~\cite{Bohr1}:
\begin{equation}
V(r) = - G {m_1 m_2 \over r}\left[1+3 \frac{G (m_1+m_2)}{c^2r}+
         \frac{41}{10\pi} \frac{G \hbar}{c^3 r^2}\right]
\end{equation}
In the above expressions we have restored the appropriate physical factors $c$ and $\hbar$.

On the grounds of dimensional analysis alone one can anticipate this form of the lowest-order corrections
to the Newtonian potential~\cite{Donoghue1}. The relativistic classical corrections are proportional to
$\ell_{cl}/r$, where
\begin{equation}
\ell_{cl}=\frac{G m}{c^2}
\end{equation}
\noindent is the classical length for the mass $m$, and the quantum
corrections (also relativistic) are proportional to
${\ell_p}^2/{r^2}$, where
\begin{equation}
\ell_p = \sqrt{\frac {G \hbar}{c^3}}
\end{equation}
\noindent $\ell_p=1.6 \times 10^{-35}$ m is the Planck length.

The classical and quantum pieces of (43) arise from the same loop diagrams, and
the order of magnitude of the quantum corrections ${G \hbar}/{c^3 r^2}$ can be derived from
the classical one using~\cite{Holstein1} the concept of ``zitterbewegung''. In fact, in transition
from classical to quantum corrections the classical distance $r$ between two masses $m_1$ and $m_2$
must be modified by an uncertainty of the order the Compton wavelengths of each masses:
\begin{equation}
r \rightarrow r + \frac{\hbar}{m_1 c} + \frac{\hbar}{m_2 c}
\end{equation}
\noindent and
\begin{equation}
\frac{1}{r} \rightarrow \frac{1}{r} - \frac{\hbar}{(m_1+m_2) c r^2} + \ldots ,
\end{equation}
\noindent so that the quantum corrections of can be understood as ``zitterbewegung'' effects applied
to the classical distance $r$.

It should be noticed that the classical post-Newtonian term in the expression (43) corresponds to the
lowest-order scattering potential and agrees with Eq. 2.5 of Iwasaki~\cite{Iwasaki}.
The correct result for the quantum corrections first published in~\cite{Bohr1} and later was confirmed
in~\cite{Khr2}.

\section{Conclusion and outlook}

The result (41) for the leading quantum corrections to the Newton law could be written in the form:
\begin{equation}
V(r) = - G {m_1 m_2 \over r}\left[1+\delta_{QC}\right]
\end{equation}
\noindent where
\begin{equation}
\delta_{QC} = \frac{41}{10\pi} \frac{{\ell_p}^2}{r^2}
\end{equation}

There are also additional quantum corrections due to the
contributions to the vacuum polarization by photons and massless neutrinos:

\begin{figure}[ht]
\begin{minipage}[t]{0.46\linewidth}
\begin{center}
\includegraphics[scale=1]{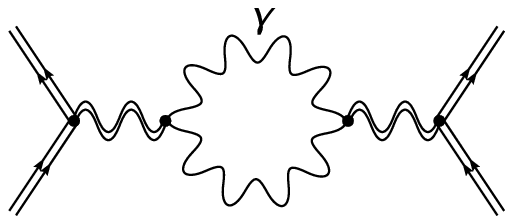}
\end{center}
\end{minipage}
\begin{minipage}{0.1\linewidth}\ \end{minipage}
\begin{minipage}[t]{0.46\linewidth}
\begin{center}
\includegraphics[scale=1]{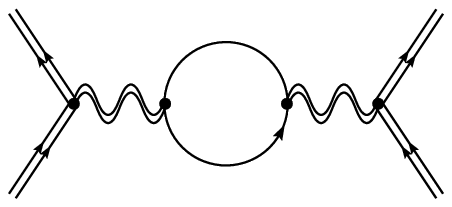}
\end{center}
\end{minipage}
\end{figure}

\noindent which were calculated by Radkowski~\cite{Rad}, Capper,
Duff, and Halpern~\cite{Capp2}, Capper and Duff~\cite{Capp3}, Duff
and Liu~\cite{Duff2}:

\begin{equation}
{\delta_{QC}}^{\gamma\nu} = \frac{4+N_{\nu}}{15\pi} \frac{{\ell_p}^2}{r^2}
\end{equation}
\noindent where $N_\nu$ is the number of massless two-component neutrinos.

The value of the both quantum corrections are controlled by the Planck length $\ell_p$,
the corrections vanish at large values of $r$ and it is accompanied by a very small coefficient,
so even for astronomical purposes these corrections are irrelevant and unlikely
to be measured in the foreseeable future. Nevertheless, such predictions would need to be
replicated by any candidate theory of high energy quantum gravity.

Only at $r_0\simeq \ell_p$ the quantum corrections become large.
But in this regime the effective field theory approach breaks down.

However, from a cosmological view point there is a cumulative effect of gravity and,
given a fixed density of energy, the integration of this effect over large volumes could give an observable
signal~\cite{Espriu}. The authors of ref.~\cite{Espriu} have found that during inflation,
the quantum corrections are significant, leading to deviations from the standard inflationary expansion.


\section{Acknowledgments}

We thank P.~R.~Anderson and A.~Fabbri for discussion. We are
grateful to Professor Jos\'e Bernab\'eu  for support and numerous
helpful remarks. A.~A. would like to thank the DGITT of the
Generalitat Valenciana for the Grant AINV06/008 and the Theoretical
Physics Department of the University of Valencia, the hospitality
extended to him during the course of this work. A.~A. thanks also
the Abdus Salam International Centre for Theoretical Physics for the
financial support in the initial stage of this project.


\begin{thebibliography}{99}

\bibitem{Einstein} A.~Einstein, {\it Ann. der Phys.} {\bf 49}, 769 (1916).

\bibitem{Dewitt} B.~S.~Dewitt, {\it Phys. Rev.} {\bf 160}, 1113 (1967);
 {\bf 162}, 1195 (1967); {\bf 162} 1239 (1967).

\bibitem{Gupta}
S.~N.~Gupta, {\it Proc. Phys. Soc.} {\bf A65}, 608 (1952);\\
S.~N.~Gupta and S.~F.~Radford, {\it Phys. Rev.} {\bf D21}, 2213
(1980).

\bibitem{Faddeev} L.~D.~Faddeev and V.~N.~Popov, {\it Phys. Lett.} {\bf 25B}, 29 (1967).

\bibitem{Mandelstam} S.~Mandelstam, {\it Phys. Rev} {\bf 175}, 1604 (1968).

\bibitem{Schwinger1} J.~Schwinger, {\it Phys. Rev.} {\bf 130}, 1253 (1963).

\bibitem{Schwinger2} J.~Schwinger, {\it Phys. Rev.} {\bf 173}, 1264 (1968).

\bibitem{Weinberg} S.~Weinberg, {\it Phys. Rev.} {\bf 135}, B1049 (1964).

\bibitem{Iwasaki} Y.~Iwasaki, {\it Prog. Theor. Phys.} {\bf 46}, 1587 (1971).

\bibitem{Veltman1} G.~'t Hooft and M.~J.~G.~Veltman, {\it Annales Poincare Phys.
Theor.} {\bf A20}, 69 (1974).

\bibitem{Veltman2} M.~Veltman, Gravitation, in {\it Les Houches, Session XXVIII, 1975},
eds. R.~Balian and J.~Zinn-Justin, (North-Holland, Amsterdam,1976),
p. 266.

\bibitem{Rad} A.~F.~Radkowski, {\it Ann. of Phys.} {\bf 56}, 319 (1970).

\bibitem{Capp1} D.~M.~Capper, G.~Leibbrandt and M.~Ramon Medrano,
{\it Phys. Rev.} {\bf D8}, 4320 (1973).

\bibitem{Duff1} M.~J.~Duff, {\it Phys. Rev.} {\bf D9}, 1837 (1974).

\bibitem{Capp2} D.~M.~Capper, M.~J.~Duff and L.~Halpern, {\it Phys. Rev.} {\bf D10}, 461 (1974).

\bibitem{Capp3} D.~M. Capper and M.~J.~Duff, {\it Nucl. Phys.} \textbf{B44}, 146 (1974).

\bibitem{Duff2} M.~J.~Duff and J.~T.~Liu, {\it Phys. Rev. Lett.} \textbf{85}, 2052 (2000).

\bibitem{Odin1} S.~D.~Odintsov and I.~L. Shapiro, {\it Class. Quant. Grav.} \textbf{9},
873 (1992).

\bibitem{Odin2} E.~Elizalde, S.~D.~Odintsov and I.~L. Shapiro, {\it Class. Quant.
Grav.} \textbf{11}, 1607 (1994).

\bibitem{Odin3} E.~Elizalde, C.~O.~Lousto, S.~D.~Odintsov and A.~Romeo,
{\it Phys. Rev.} \textbf{D52}, 2202 (1995).

\bibitem{Mazz1} D.~A.~R.~Dalvit and F.~D.~Mazzitelli, {\it Phys. Rev.} \textbf{D50}, 1001
(1994).

\bibitem{Mazz2} D.~A.~R.~Dalvit and F.~D.~Mazzitelli, {\it Phys. Rev.} \textbf{D56}, 7779 (1997).

\bibitem{Mazz3} A.~Satz, F.~D.~Mazzitelli and E.~Alvarez, {\it Phys. Rev.}
\textbf{D71}, 064001 (2005).

\bibitem{Paul} P.~R.~Anderson and A.~Fabbri, gr-qc/0612018.

\bibitem{Goroff} M.~H.~Goroff and A.~Sagnotti, {\it Nucl. Phys.} {\bf B266}, 709 (1986).

\bibitem{vandeVen} A.~E.~van de Ven, {\it Nucl. Phys.} {\bf B378}, 309 (1992).

\bibitem{Weinberg1} S.~Weinberg, in {\it General Relativity. An Einstein Centenary
Survey}, eds. S.~W.~Hawking and W.~Israel (Cambridge University
Press, Cambridge, 1979), p. 790.

\bibitem{Weinberg2} S.~Weinberg, {\it Physica} {\bf A96}, 327 (1979);\\
         J.~Gasser and H.~Leutwyler, {\it Nucl. Phys.} {\bf B250}, 465 (1985).

\bibitem{Shiekh1}
A.~Y.~Shiekh, {\it Can. J. Phys.} {\bf 74}, 172 (1996) [gr-qc/9307100];\\
A.~Y.~Shiekh, The perturbative quantization of gravity, in {\it
Proc. XVII Int. Workshop on Problems of High Energy physics and
Field Theory}, eds. A.~P.~Samokhin, G.~L. Rcheulishvili ( IHEP,
Protvino, 1995), p. 156 [hep-th/9407159].

\bibitem{Donoghue1}
  J.~F.~Donoghue, {\it Phys. Rev. Lett.} {\bf 72}, 2996 (1994)[gr-qc/9310024];\\
  J.~F.~Donoghue, {\it Phys. Rev.} {\bf D50}, 3874 (1994) [gr-qc/9405057].

\bibitem{Deser} S.~Deser and P.~van~Nieuwenhuizen,
{\it Phys. Rev. Lett.} {\bf 32}, 245 (1974);\\
{\it Phys. Rev.} {\bf D10}, 401 (1974); {\it Phys. Rev.} {\bf D10},
411 (1974).

\bibitem{Gast} F.~Berends and R.~Gastmans, {\it Phys. Lett.} {\bf B55}, 311
(1975).

\bibitem{Abbott} L.~F.~Abbott, {\it Nucl. Phys.} {\bf B185}, 189 (1981).

\bibitem{Muzinich}
I.~J.~Muzinich and S.~Vokos, {\it Phys. Rev.} {\bf D52}, 3472
(1995).

\bibitem{Hamber}
H.~W.~Hamber and S.~Liu, {\it Phys. Lett.} {\bf B357}, 51 (1995).

\bibitem{Akh} A.~A.~Akhundov, S.~Bellucci and A.~Shiekh,
{\it Phys. Lett.} {\bf B395}, 16 (1997).

\bibitem{Khr1} I.~B.~Khriplovich and G.~G.~Kirilin,
{\it J. Exp. Theor. Phys.} {\bf 95}, 981 (2002); {\it Zh. Eksp.
Teor. Fiz.} {\bf 95}, 1139 (2002).

\bibitem{Bohr1}
  N.~E.~J.~Bjerrum-Bohr, J.~F.~Donoghue and B.~R.~Holstein,
  {\it Phys. Rev.} {\bf D67}, 084033 (2003)
  [{\it Erratum-ibid.} {\bf D71}, 069903 (2005)]

\bibitem{Khr2} I.~B.~Khriplovich and G.~G.~Kirilin,
{\it J. Exp. Theor. Phys.} {\bf 98}, 1063 (2004); {\it Zh. Eksp.
Teor. Fiz.} {\bf 125}, 1219 (2004).

\bibitem{Kir1} G.~G.~Kirilin, {\it Nucl.Phys.} {\bf B728}, 179 (2005).

\bibitem{Ein} A.~Einstein, L.~Infeld and B.~Hoffmann, Ann. Math. \textbf{39}, 65
(1938).
\bibitem{Edd} A.~Eddington and G.~Clark, {\it Proc. Roy. Soc.} \textbf{166}, 465 (1938).

\bibitem{Okamura}
K.~Hiida and H.~Okamura, {\it Prog. Theor. Phys.} {\bf 47}, 1743
(1972).

\bibitem{Barker:bx}
B.~M.~Barker and R.~F.~O'Connell, {\it J. Math. Phys.} {\bf 18},
1818 (1977) [{\it Erratum-ibid.} {\bf 19}, 1231 (1978)].

\bibitem{Barker:ae}
B.~M.~Barker and R.~F.~O'Connell, {\it Phys. Rev.} {\bf D12}, 329
(1975).

\bibitem{Bohr2}
N.~E.~J.~Bjerrum-Bohr, J.~F.~Donoghue and B.~R.~Holstein, {\it Phys.
Rev.} {\bf D68}, 084005 (2003) [{\it Erratum-ibid.} {\bf D71},
069904 (2005)].

\bibitem{Kir2} G.~G.~Kirilin, gr-qc/0601020.

\bibitem{Bohr} N.~E.~J.~Bjerrum-Bohr,
{\it Phys. Rev.} {\bf D66} 084023 (2002).

\bibitem{Butt}
M.~S.~Butt, gr-qc/0605137.

\bibitem{Burgess}
C.~P.~Burgess, {\it Liv. Rev. Rel.} {\bf 7}, 5 (2004);\\
C.~P.~Burgess, gr-qc/0606108.

\bibitem{Holstein}
B.~R.~Holstein, gr-qc/0607045.

\bibitem{Kazakov}
K.~A.~Kazakov, {\it Phys. Rev.} {\bf D63}, 044004 (2001).

\bibitem{Lee} J.M. Lee, {\it Ricci, A Mathematica package for doing tensor
calculations in differential geometry}, (User's Manual, Version 1.2,
1992-1995).

\bibitem{Math} S. Wolfram, {\it Mathematica} (Addison-Wesley, Reading, MA, 1991).

\bibitem{JaxoDraw} D.~Binisi and L.~Theussl, {\it Comp. Phys. Comm.} {\bf 161}, 76 (2004).

\bibitem{Holstein1}
B.~R.~Holstein, {\it Eur. Phys. J.} {\bf A18}, 227 (2003).

\bibitem{Espriu} D.~Espriu, T.~Multamaki and E.~C.~Vagenas,
{\it Phys. Lett.} \textbf{B628}, 197 (2005).

\end{thebibliography}
\end{document}